# Tailoring Semiconductor Lateral Multi-junctions for Giant Photoconductivity Enhancement

Yutsung Tsai[1†], Zhaodong Chu[1†], Yimo Han[2], Chih-Piao Chuu[3,4], Di Wu[1], Alex Johnson[1], Fei Cheng[1], Mei-Yin Chou[3,5], David A. Muller[2,6], Xiaoqin Li[1], Keji Lai[1]*, Chih-Kang Shih[1]*

[1]*Department of Physics, Center for Complex Quantum Systems, The University of Texas at Austin, Austin, Texas 78712, USA*

[2]*School of Applied and Engineering Physics, Cornell University Ithaca, NY 14853, USA*

[3]*Institute of Atomic and Molecular Sciences, Academia Sinica, Taipei 10617, Taiwan*

[4]*Physics Division, National Center for Theoretical Sciences, Hsinchu, Taiwan*

[5]*School of Physics, Georgia Institute of Technology, Atlanta, GA 30332, USA*

[6]*Kavli Institute at Cornell for Nanoscale Science, Ithaca, New York 14853, USA*

[†] These authors contributed equally to this work

* E-mails: kejilai@physics.utexas.edu and shih@physics.utexas.edu

# Abstract

Semiconductor heterostructures have played a critical role as the enabler for new science and technology. The emergence of transition metal dichalcogenides (TMDs) as atomically thin semiconductors has opened new frontiers in semiconductor heterostructures either by stacking different TMDs to form vertical heterojunctions or by stitching them laterally to form lateral heterojunctions via direct growth. In conventional semiconductor heterostructures, the design of multi-junctions is critical to achieve carrier confinement. Analogously, we report successful synthesis of monolayer $WS_2/WS_{2(1-x)}Se_{2x}/WS_2$ multi-junction lateral heterostructure via direct growth by chemical vapor deposition. The grown structures are characterized by Raman, photoluminescence, and annular dark-field scanning transmission electron microscopy to determine its lateral compositional profile. More importantly, using microwave impedance microscopy, we demonstrate that the local photoconductivity in the alloy region can be tailored and enhanced by 2 orders of magnitude over pure $WS_2$. Finite element analysis confirms that this effect is due to the carrier diffusion and confinement into the alloy region. Our work exemplifies the technological potential of atomically thin lateral heterostructures in optoelectronic applications.

Inspired by the widespread applications of semiconductor heterostructures in modern transistors[1,2], photodetectors[3,4] and solid-state lasers[5], vertically stacked two-dimensional (2D) van der Waals (vdW) materials[6] have attracted substantial research interest, where many exciting phenomena such as interlayer excitons[7,8], photovoltaic effect[9], tunneling assisted carrier recombination[10], and negative differential resistance[11,12] have been reported. On the other hand, 2D materials offer another unique opportunity of forming lateral hetero-junctions within the same atomic layer[13]. To date, single lateral junctions of transition metal dichalcogenide (TMD) monolayers with different cation or anion configurations have been demonstrated, showing the desirable p-n diode effect and enhanced photoluminescence[13-16]. Going beyond a single TMD lateral junction, multiple in-plane junctions can offer greater flexibility in device design.

In this work, we report the epitaxial growth of a lateral $WS_2/WS_{2(1-x)}Se_{2x}/WS_2$ ($x \approx 0.15$) multi-junction and the characterization of its properties by various microscopy techniques. Intriguingly, the alloy region shows a giant photoconductivity that is 2 orders of magnitude higher than that of the $WS_2$ under light illumination. Local gating experiment and finite-element modeling suggest that the enhancement of photo-response is due to the accumulation of electrons in the alloy section with a sub-micrometer lateral width. Our work represents an important step toward engineering lateral multi-junctions for optoelectronic devices with novel functionalities.

Unlike vertical heterostructures, where preceding layers are naturally protected in the subsequent growth, vdW materials with lateral junctions are fully exposed in the chemical environment during the synthesis process. Therefore, optimal conditions have to be met to ensure activated edge growth [17-22]. The lateral multi-junction sample (Sample A) is grown by an *in situ* 3-step process in the chemical vapor deposition (CVD) chamber. Details of the CVD setup are shown in Supporting Information Figure S1. In the first step, the sulfur vapor in the $H_2$/Ar carrier gas is introduced to the central furnace to react with the $WO_3$ powder, resulting in the growth of $WS_2$ crystals on the double-side-polished (DSP) sapphire substrates. In the second step, only the heating tape surrounding the selenium precursor is turned on, introducing selenium vapor to the upstream of the furnace. We find that under certain conditions (see Methods), the $WS_{2(1-x)}Se_{2x}$ alloy is formed at the activated $WS_2$ edge involving Se-S exchange replacement, leading to mostly inward and some outward growth on the monolayer flakes. In the last step, sulfur vapor is again fed into the central furnace for the growth of the outermost $WS_2$ triangular ring. The optical image of a typical flake in Sample A is shown in Figure 1a. As a control experiment, we have also grown a reference sample (Sample B) without Step 3, resulting in a single $WS_2$/alloy junction rather than the $WS_2$/alloy/$WS_2$ multi-junctions.

To evaluate the Se composition $x$ in the alloy region, we have performed atomically resolved annular dark-field scanning transmission electron microscopy (ADF-STEM) on our samples (Supporting Information S4)[14,23,24]. As shown in Figure 1b, the bright W atoms are arranged in a triangular lattice in all three domains. Minimum lattice distortion of the alloy is observed from the single-point diffraction pattern, indicating that its average lattice constant is similar to that of the $WS_2$. In the alloy region, three configurations with different contrast in the center of the W triangles can be seen, i.e., [S, S], [S, Se], and [Se, Se] vertical stacks with negligible, weak, and strong contrast, respectively. Within the 15 nm × 15 nm scan range, the average Se doping is estimated to be $x = 13\%$ by counting the number of Se sites.

Figures 1c and 1d show the confocal micro-Raman spectroscopy data on the same flake in Figure 1a. Due to the Se substitution, one of the out-of-plane $A_{1g}$ modes shifts from 422 cm$^{-1}$ in $WS_2$ to 417 cm$^{-1}$ in the $WS_{2(1-x)}Se_{2x}$ alloy. The other $A_{1g}$ mode at 267 cm$^{-1}$, commonly observed in $WSe_2$ but absent in $WS_2$, appears only in the alloy region[21,25-28]. The Raman modes of $WS_2$ and alloy are consistent with our theory calculation (Supporting Information S3). The integrated Raman maps around 422 cm$^{-1}$ and 267 cm$^{-1}$ in Figure 1c clearly display the lateral $WS_2$/$WS_{2(1-x)}Se_{2x}$/$WS_2$ heterostructure of Sample A, whereas the $WS_2$/alloy core-shell structure is seen in Raman mapping for Sample B (Supporting Information S4). In addition to the shift of the Raman spectrum, the peak wavelength of the photoluminescence (PL) signals also moves from 627 nm in pure $WS_2$ to 643 nm in the alloy region shown in Figure 1e and 1f. Given that the PL of $WSe_2$ peaks at 750 nm, we can again estimate the Se composition $x \approx 15\%$[21], which is in good agreement with the ADF-STEM results.

In order to study the photo-response of multi-junction TMD samples, we perform local photoconductivity imaging using light-stimulated microwave impedance microscopy (MIM) [29,30], as illustrated in Figure 2a. Here, the sample is illuminated by the focused 532-nm laser beam from below the sapphire substrate. During the experiment, the sample stage is scanned in the $xy$-plane, whereas the center of the laser spot (~ 4 µm in diameter) is always aligned with the MIM cantilever probe mounted on the *z*-scanner. Unlike the widely used scanning photocurrent microscopy (SPCM) [31, 32], which detects the photocurrent across micro-fabricated source and drain electrodes, the MIM measures the imaginary (MIM-Im) and real (MIM-Re) parts of the tip-sample admittance, from which the local conductivity of the sample can be extracted. Figure 2b shows the finite-element analysis (FEA) of the MIM response using the actual tip/sample geometry and material properties. A 15 nm $Al_2O_3$ protection layer on top of the TMD is taken into account explicitly. Details of the FEA modeling are described in Supporting Information S5. As more carriers are injected to the TMD layer with an increasing laser power, the microwave electric fields become more confined between the tip and the 2D material due to the screening effect. As a result, the MIM-Im signal (proportional to the tip-sample capacitance) increases monotonically as a function of the TMD conductivity $\sigma_{2D}$ and saturates for $\sigma_{2D} > 10^{-4}$ S·sq. On the other hand, the MIM-Re signal (proportional to the electrical loss), peaks around $\sigma_{2D} = 10^{-6}$ S·sq and decreases for both higher and lower $\sigma_{2D}$. The field profile in Figure 2c also shows that the spatial resolution of the MIM is comparable to the tip diameter of about 200 nm.

The MIM images of a typical $WS_2$/alloy/$WS_2$ flake at selected laser powers (*P*) are displayed in Figure 2d. It is worth noting that the multi-junction samples are morphologically stable under illumination, as seen by the same AFM data during the light-stimulated MIM measurements. A complete set of the MIM images and a video clip of the data are included in Supporting Information S6. Strikingly, much stronger photo-induced MIM signals are measured in the alloy region than that in the inner and outer $WS_2$ domains. We caution that the apparent width $w_{alloy}$ of 0.2 μm in the MIM-Im data may be limited by the spatial resolution. The same phenomenon has been observed in other flakes in Sample A (Supporting Information S7). In Figure 3a, the averaged MIM signals in both the alloy and $WS_2$ domains are plotted as a function of *P*. By comparing with the simulated response curves in Figure 2b, the MIM signals at these two regions can be converted to the local photoconductivity, as shown in Figure 3b. Assuming that the *n*-type photo-generated carriers are responsible for the transport, whereas the low-mobility *p*-type carriers are mostly trapped, for a 2D semiconductor with incident photon-to-electron conversion efficiency (IPCE) $\eta$, carrier recombination time $\tau$, and carrier mobility $\mu$, the 2D photoconductivity under above-gap (photon energy $h\nu$, elementary charge $e$) illumination scales with the power intensity *P* as follows.

$$\sigma_{2D} = \eta \cdot \frac{P\tau}{h\nu} \cdot e \cdot \mu = n \cdot e \cdot \mu \tag{1}$$

Taking order-of-magnitude values for the estimate, such as $\eta \approx 0.1\%$ [33], $\tau \approx 0.1$ μs [34], $\mu \approx 10$ cm$^2$ V$^{-1}$ s$^{-1}$ [34-36], and $h\nu$ = 2.33 eV, the linear $\sigma_{2D} \propto P$ relation in the WS$_2$ region is in good agreement with that of the monolayer TMD materials reported in the literature [34]. The much (~ 100 times) higher $\sigma_{2D}$ in the alloy region, on the other hand, can only be explained by Eq. (1) with the much larger carrier density $n$. To exclude the possibility that the giant photo-response is merely a result of the material property of WS$_{2(1-x)}$Se$_{2x}$, we have also performed the light-stimulated MIM experiment on Sample B. As seen in Figures 3b and 3c, while WS$_2$ shows the same MIM signals in both samples, the alloy section in the single-junction Sample B exhibits a photoconductivity that is ~ 3 times lower than that of WS$_2$, presumably due to the alloy scattering effect. The remarkably high photo-response in the alloy region of Sample A is therefore specific to the WS$_2$/WS$_{2(1-x)}$Se$_{2x}$/WS$_2$ multi-junction heterostructure.

In order to understand the 100-fold enhancement of photoconductivity in the alloy region of Sample A, we model the carrier diffusion process by the rate equation[37,38],

$$\frac{\partial n}{\partial t} = \varphi - \frac{n - n_0}{\tau} + D\nabla^2 n \tag{2}$$

where $n$ is the local carrier concentration, $\varphi = \eta P/h\nu$ is the photo-carrier generation rate, $n_0 \approx 0$ is the carrier concentration under thermal equilibrium, $D$ is the diffusion constant, and $L = \sqrt{D\tau}$ is the diffusion length. In the steady state with $\partial n/\partial t = 0$, the equilibrium carrier distribution can be simulated by the FEA. Note that in our experiment, the MIM only measures the photo-response at the center of the laser spot, which is modeled as a uniformly illuminated disk with a diameter $d$ = 4 μm. To capture the effect of charge accumulation in WS$_{2(1-x)}$Se$_{2x}$, we assume that the carriers diffused to the alloy region are all confined here, as illustrated by the simple 1D model in Figure 3d. In the limit of large $L > d$, half of the photo-generated carriers will flow to the alloy, resulting in an enhanced carrier density $n_{alloy}$ that is a factor of $d/2w_{alloy}$ higher than the photo doping of $\varphi\tau$. The confinement effect does not occur in the WS$_2$ region. In the limit of large $L$, the photo-generated carriers will be evenly distributed in the entire flake, resulting in a carrier density $n_{WS2} \approx (d/w)^2 \cdot \varphi\tau$, where $w$ = 6 μm is the lateral size of the sample. The dependence of $n_{alloy}/n_{WS2}$ as a function of $L$ is summarized in Figure 3e. The ratio starts from 1 for small $L$, increases rapidly once $L > w_{alloy}$, and saturates at ~ 20 for $L > 10$ μm. As is shown in the inset of Figure 3e, the enhancement effect does not depend on the laser power, which is consistent with our experimental observations. Note that $w_{alloy} = 0.2$ μm in the calculation is based on the MIM data, which may overestimate the actual dimension of the lateral confinement. In addition, in the presence of concentration gradient, the effective confinement width would be smaller than the actual width of the alloy region, A smaller effective confinement width of 50 nm and a diffusion length of $L \approx 10$ μm will lead to $n_{alloy}/n_{WS2} \approx 100$ that is consistent with the experimental data. In all, our analysis highlights the importance of carrier diffusion and confinement effects in the lateral multi-junction TMD structures.

Finally, we briefly comment on the origin of the lateral confinement effect. Figure S8a shows the schematic of light-stimulated MIM measurement with a DC tip bias $V_{tip}$ on a multi-junction flake contacted by a Ti(20 nm)/Pd(20 nm) electrode. As is seen in Figure S8b, the MIM response is weaker when the tip is biased negatively with respect to the sample and stronger for a positive $V_{tip}$, which indicates that the majority carriers responsible for photoconductivity are *electrons*. In Supporting Information S9, we include the density-functional theory (DFT) calculation of the band structure of $WS_2$ and $WS_{1.69}Se_{0.31}$ ($x = 15.5\%$). A type-II band alignment is predicted for the interface with a symmetric arrangement of Se in the alloy, resulting in a smaller conduction band offset (50 meV) than the valence band offset (120 meV). The trend is consistent with the relative alignment between the energy bands of $WS_2$ and $WSe_2$.[39] This type-II band alignment cannot explain the trapping of electron carriers found in the alloy region. However, we note that our ADF-STEM data show more [S, Se] than [Se, Se] vertical stacks in the alloy region, which is understandable because the Se vapor comes from one side of the monolayer in the replacement of S. This Se distribution pattern gives rise to a small vertical dipole in the alloy region, which in turn shifts the energy bands. The shifted band alignment due to this vertical dipole in the alloy region would be a topic of great interest for future experimental and theoretical studies.

In summary, we have demonstrated a remarkably enhancement of local photoconductivity in a lateral TMD heterostructure grown via three-step *in-situ* CVD synthesis for the first time. The core-ring-ring structure of $WS_2$/ $WS_{1.75}Se_{0.25}$/$WS_2$ configuration is confirmed by Raman, PL and ADF-STEM measurements. Giant photoconductivity measured by light-stimulated MIM measurement in the alloy domain verified the desired functionality of the carrier confinement in the TMD lateral heterojunctions. By achieving the critical step of synthesizing multi-junction lateral heterostructure and demonstrating the carrier confinement effect, our work significantly expands the TMD lateral heterostructures to a more complex architecture promising for advanced optoelectronic devices.

## Methods

**Synthesis Conditions for CVD Growth**

As shown in Figure S1a, we used $H_2$ and Ar carrier gases at ambient pressure with a $H_2$/Ar flow rate of 7/70 sccm. The chalcogenide precursors, the sulfur powder, and the selenium powder, were placed in alumina boats outside of the central furnace at the upstream end of the 1-inch quartz tube. Selenium powder was closer to the furnace center with heating tape I whereas sulfur powder was further away from the center of the furnace with heating tape II; the $WO_3$ powder 0.4-0.5g was kept in the center hot zone during the whole growth; sapphire substrates were suspended above an alumina slide fitted in a 0.7-inch quartz tube. $WS_2$ crystals were grown for 10 mins and naturally cooled down to 750℃ before the second growth step. $WS_{2(1-x)}Se_{2x}$ alloy was grown for 10 mins at the activated $WS_2$ edge and naturally cooled down to 750℃ before the third growth step. $WS_2$ crystals at the activated $WS_{2(1-x)}Se_{2x}$ edge were grown for 10 mins and naturally cooled down to room temperature till the end of the growth. We kept the ambient pressure during the whole growth period without breaking the vacuum to ensure the edge of each stage remained active for the sequential epitaxial growth.

**Characterization**

The lateral multi-junction samples were characterized by optical microscope, micro-Raman, micro-PL, and ADF-STEM. The composite micro-Raman and micro-PL were conducted using the confocal Raman system Witec Micro-Raman Spectrometer Alpha 300 with Ar ion laser (488nm) excitation. No change to the Raman or PL response was observed several months after the growth, indicating that the samples are stable under the ambient condition. The multi-junction devices with top-contact electrodes were fabricated using electron-beam lithography followed by electron-beam deposition of metal thin films. The ADF-STEM images were acquired using a FEI Titan Themis CryoS/TEM at 80 kV with 10 pA beam current.

**Associated content**

Supplementary Information.

**Acknowledgment**

The work at UT-Austin led by X.L., K.L., C.K.S. and M.Y.C. is supported by NSF EFMA-1542747. Y.T, A.J., F.C and C.K.S. acknowledge Welch Foundation Grant No. F-1672. Z.C and D.W. acknowledge the support from Welch Foundation Grant No. F-1814. Raman and PL work is conducted in Texas Material Institute at UT Austin. Electron microscopy work used the electron microscopy facilities supported by the Cornell Center for Materials Research, and NSF MRSEC (DMR-1120296). C.P.C. and M.Y.C. acknowledge the support from a Thematic Project at Academia Sinica.

**References**

[1] H. Morkoc, S. Strite, G.B. Gao, M.E. Lin, B. Sverdlov, M.Burns, J. Appl. Phys. **1994**, 76, 1363.

[2] J. Xiang, W. Lu, Y. Hu, Y. Wu, H. Yan, C. M. Lieber, Nature **2006**, 441, 489.

[3] B.F. Levine, J. Appl. Phys. **1993**, 74, R1.

[4] M. Massicotte, P. Schmidt, F. Vialla, K. G. Schädler, A. Reserbat-Plantey, K. Watanabe, T. Taniguchi, K. J. Tielrooij, F. H. L. Koppens, Nat. Nanotechnol. **2015**, 11, 42.

[5] J. Faist, F. Capasso, D.L. Sivco, C. Sirtori, A.L. Hutchinson, A.Y. Cho, Science **1994**, 264, 553.

[6] D. Jariwala, T.J. Marks, M.C. Hersam, Nat. Mater. **2016**, 16, 170.

[7] P. Rivera, J. R. Schaibley, A. M. Jones, J. S. Ross, S. Wu, G. Aivazian, P. Klement, K. Seyler, G. Clark, N. J. Ghimire, J. Yan, D. G. Mandrus, W. Yao, X. Xu, Nat. Commun. **2015,** 6, 6242.

[8] H. Chen, X. Wen, J. Zhang, T. Wu, Y. Gong, X. Zhang, J. Yuan, C. Yi, J. Lou, P. M. Ajayan, W. Zhuang, G. Zhang, J. Zheng, Nat. Commun. **2016** 7, 12512.

[9] M. M. Furchi, A. Pospischil, F. Libisch, J. Burgdorfer, T. Mueller, Nano Lett. **2014**, 14, 4785.

[10] C.-H. Lee, G.-H. Lee, A. M. van der Zande, W. Chen, Y. Li, M. Han, X. Cui, G. Arefe, C. Nuckolls, T. F. Heinz, J. Guo, J. Hone, P. Kim. Nat, Nanotechnol. **2014**, 9, 676.

[11] Y.-C. Lin, R. K. Ghosh, R. Addou, N. Lu, S. M. Eichfeld, H. Zhu, M.-Y. Li, X. Peng, M. J. Kim, L.-J. Li, R. M. Wallace, S. Datta, J. A. Robinson, Nat. Commun. **2015,** 6, 7311.

[12] J. Shim, S. Oh, D.-H. Kang, S.-H. Jo, M. H. Ali, W.-Y. Choi, K. Heo, J. Jeon, S. Lee, M. Kim, Y. J. Song, J.-H. Park, Nat. Commun. **2016,** 7, 13413.

[13] M.-Y. Li, Y. Shi, C.-C. Cheng, L.-S. Lu, Y.-C. Lin, H.-L. Tang, M.-L. Tsai, C.-W. Chu, K.-H. Wei, J.-H. He, W.-H. Chang, K. Suenaga, L.-J. Li, Science **2015**, 349, 524.

[14] Y. Gong, J. Lin, X. Wang, G. Shi, S. Lei, Z. Lin, X. Zou, G. Ye, R. Vajtai, B. I. Yakobson, H. Terrones, M. Terrones, B. K. Tay, J. Lou, S. T. Pantelides, Z. Liu, W. Zhou, P. M. Ajayan, Nat. Mater. **2014**, 13, 1135.

[15] K. Bogaert, S. Liu, J. Chesin, D. Titow, S. Gradečak, S. Garaj, Nano Lett. **2016**, 16, 5129.

[16] J. S. Ross, P. Klement, A. M. Jones, N. J. Ghimire, J. Yan, D. G. Mandrus, T. Taniguchi, K. Watanabe, K. Kitamura, W. Yao, D. H. Cobden, X. Xu, Nat. Nanotechnol. **2014**, 9, 268.

[17] S. Zheng, L. Sun, T. Yin, A. M. Dubrovkin, F. Liu, Z. Liu, Z. X. Shen, H. J. Fan, Appl. Phys. Lett. **2015**, 106, 6.

[18] S. Tongay, D. S. Narang, J. Kang, W. Fan, C. Ko, A. V. Luce, K. X. Wang, J. Suh, K. D. Patel, V. M. Pathak, J. Li, J. Wu, Appl. Phys. Lett. **2014**, 104, 1.

[19] H. Liu, K. K. A. Antwi, S. Chua, D. Chia, Nanoscale **2014**, 6, 624.

[20] H. Li, X. Duan, X. Wu, X. Zhuang, H. Zhou, Q. Zhang, X. Zhu, W. Hu, P. Ren, P. Guo, L. Ma, X. Fan, X. Wang, J. Xu, A. Pan, X. Duan, J. Am. Chem. Soc. **2015**, 137, 5284.

[21] X. Duan, C. Wang, Z. Fan, G. Hao, L. Kou, U. Halim, H. Li, X. Wu, Y. Wang, J. Jiang, A. Pan, Y. Huang, R. Yu, X. Duan, Nano Lett. **2016**, 16, 264.

[22] J. Mann, Q. Ma, P. M. Odenthal, M. Isarraraz, D. Le, E. Preciado, D. Barroso, K. Yamaguchi, G. von S. Palacio, A. Nguyen, T. Tran, M. Wurch, A. Nguyen, V. Klee, S. Bobek, D. Sun, T. F. Heinz, T. S. Rahman, R. Kawakami, L. Bartels, Adv. Mater. **2014**, 26, 1399.

[23] R. Yan, S. Fathipour, Y. Han, B. Song, S. Xiao, M. Li, N. Ma, V. Protasenko, D. A. Muller, D. Je D, H. G. Xing, Nano Lett. **2015**, 15, 5791.

[24] Y. Han, S. Xie, B. Savitzky, R. Hovden, H. Gao, L. F. Kourkoutis, J. Park, D. A. Muller, Microsc. Microanal. **2016**, 22, 870.

[25] J. Kang, S. Tongay, J. Li, J. Wu, J. Appl. Phys. **2013**, 113, 143703.

[26] G. Gong, Z. Liu, A. R. Lupini, G. Shi, J. Lin, S. Najmaei, Z. Lin, A. L. Elías, A. Berkdemir, G. You, H. Terrones, M. Terrones, R. Vajtai, S. T. Pantelides, S. J. Pennycook, J. Lou, W. Zhou, P. M. Ajayan, Nano Lett. **2014**, 14, 442.

[27] Q. Fu, L. Yang, W. Wang, A. Han, J. Huang, P. Du, Z. Fan, J. Zhang, B. Xiang, Adv. Mater. **2015**, 27, 4732.

[28] Y. Chen, D. O. Dumcenco, Y. Zhu, X. Zhang, N. Mao, Q. Feng, M. Zhang, J. Zhang, P.-H. Tan, Y.-S. Huang, L. Xi, Nanoscale **2014**, 6, 2833.

[29] Z. Chu, M. Yang, P. Schulz, D. Wu, X. Ma, E. Seifert, L. Sun, K. Zhu, X. Li, K. Lai, arxiv.1610.00755

[30] K. Lai, W. Kundhikanjana, M. Kelly, Z. X. Shen, Rev. Sci. Instrum. **2008**, 79, 6.

[31] T. S. Kasırga, D. Sun, J. H. Park, J. M. Coy, Z. Fei, X. Xu, D. H. Cobden, Nat. Nanotechnol. **2012,** 7, 723.

[32] D. Sun, G. Aivazian, A. M. Jones, J. S. Ross, W. Yao, D. Cobden, X. Xu, Nat. Nanotechnol. **2012**, 7, 114.

[33] Y. Gong, S. Lei, G. Ye, B. Li, Y. He, K. Keyshar, X. Zhang, Q. Wang, J. Lou, Z. Liu, R. Vajtai, W. Zhou, P. M. Ajayan, Nano Lett. **2015**, 15, 6135.

[34] L. Xie, X. Cui, Proc. Natl. Acad. Sci. **2015**, 10, 1073.

[35] Y. Gong, V. Carozo, H. Li, M. Terrones, T. N. Jackson, 2D Mater. **2016,** 3, 021008.

[36] D. Ovchinnikov, A. Allain, Y. Huang, D. Dumcenco, A. Kis. ACS Nano. **2014**, 8, 8174.

[37] M. M. Furchi, D. K. Polyushkin, A. Pospischil, T. Mueller, Nano Lett. **2014**, 14, 6165.

[38] S.M. Sze, Physics of Semiconductor Devices, Wiley, USA, **1969**

[39] Y. Liang, S. Huang, R. Soklaski, L. Yang, Appl. Phys. Lett. **2013**, 103, 042106.

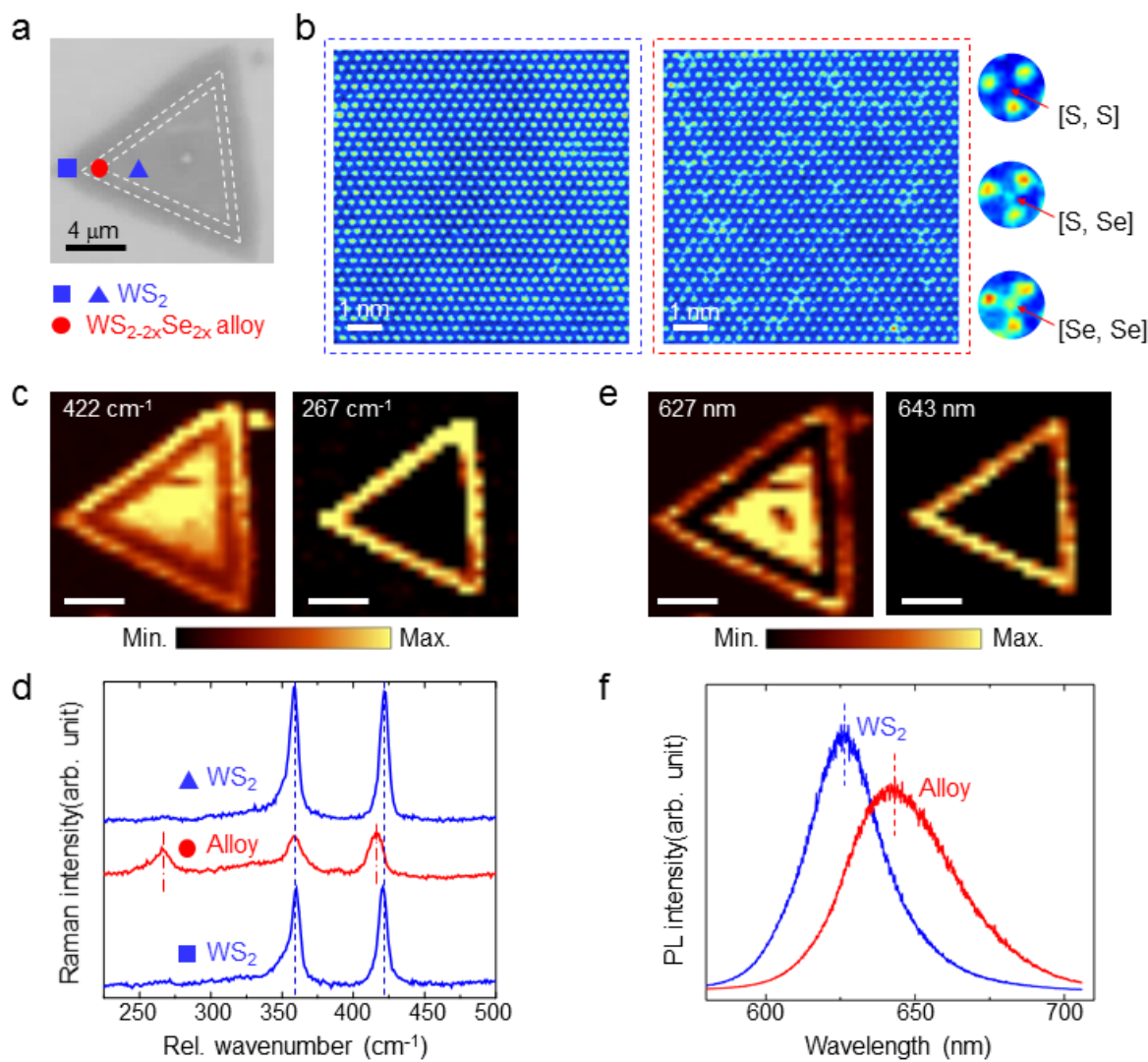

**Figure 1: Raman mapping, PL mapping and ADF-STEM scan of lateral WS$_2$/WS$_{2(1-x)}$Se$_{2x}$/WS$_2$ heterojunctions.** a) Optical image of the WS$_2$/WS$_{2(1-x)}$Se$_{2x}$/WS$_2$ heterojunction sample A b) ADF-STEM scan for the WS$_2$ domain (left) and WS$_{2(1-x)}$Se$_{2x}$ domain (right). Three configurations ([S, S], [S, Se], and [Se, Se] vertical stacks with negligible, weak, and strong contrast, respectively) in the center of the W triangles are shown to the right. c) Raman maps by integrating the WS$_2$ 422 cm$^{-1}$ mode (left) and the WS$_{2(1-x)}$Se$_{2x}$ 267 cm$^{-1}$ mode (right). d) Raman spectrum at three locations indicated in a). Scale bars in c) and d) are 4 μm. e) PL composite mapping integrated around the WS$_2$ (left, 622nm-632nm) and WS$_{2(1-x)}$Se$_{2x}$ emission peaks (right, 638-648nm). f) PL spectrum from the WS$_2$ and WS$_{2(1-x)}$Se$_{2x}$ domains, respectively.

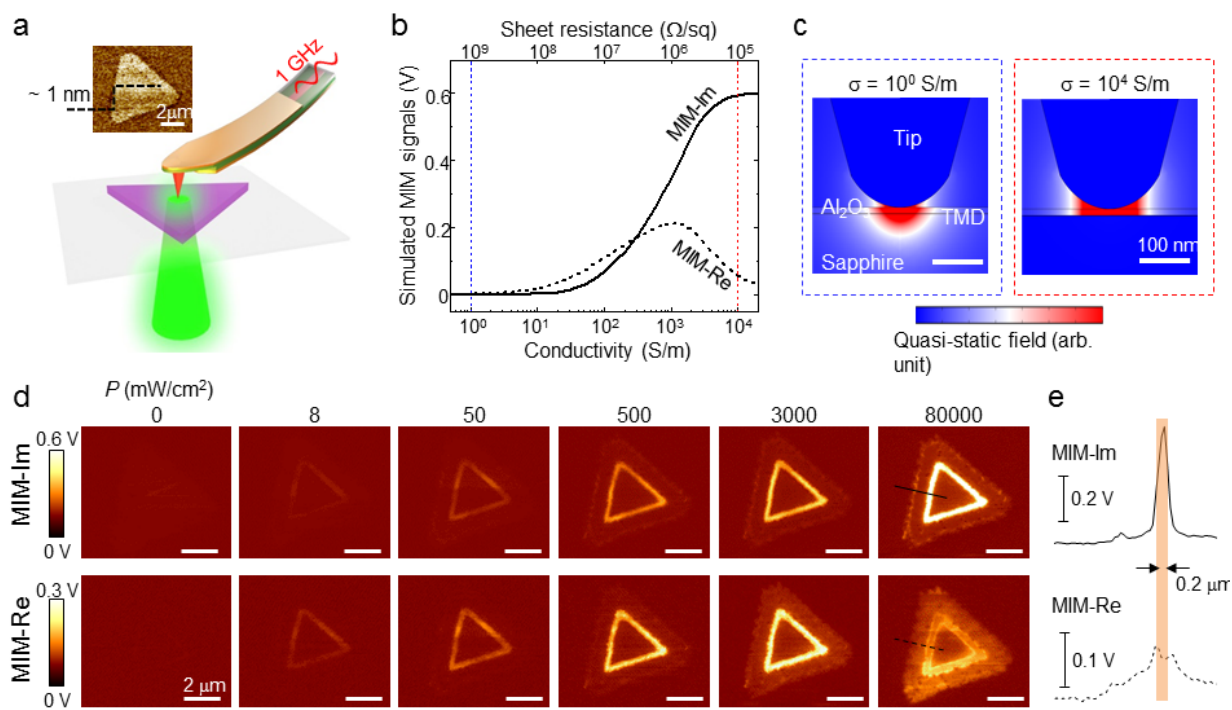

**Figure 2: Light-stimulated MIM measurements on $WS_2/WS_{2(1-x)}Se_{2x}/WS_2$.** a) Schematic diagram of the MIM setup with bottom illumination through the double polished sapphire substrate. Spatial mapping is accomplished by moving the sample stage while fixing the laser beam and the probe tip, which are aligned before scanning. Inset: AFM image of the $WS_2/WS_{2(1-x)}Se_{2x}/WS_2$ lateral multi-junction. b) Simulated MIM signals as a function of the sample conductivity $\sigma_{2D}$ and sheet resistance. c) The quasi-static displacement-field distributions at $\sigma_{2D} = 10^0$ S/m (left, blue dashed box) and $10^4$ S/m (right, red dashed box), respectively. d) Selected MIM images of the multi-junction sample as a function of the 532-nm laser intensity. All scale bars are 2 μm. e) Line profiles indicated in d), showing an enhanced MIM response in the alloy region with a width of approximately 0.2 μm.

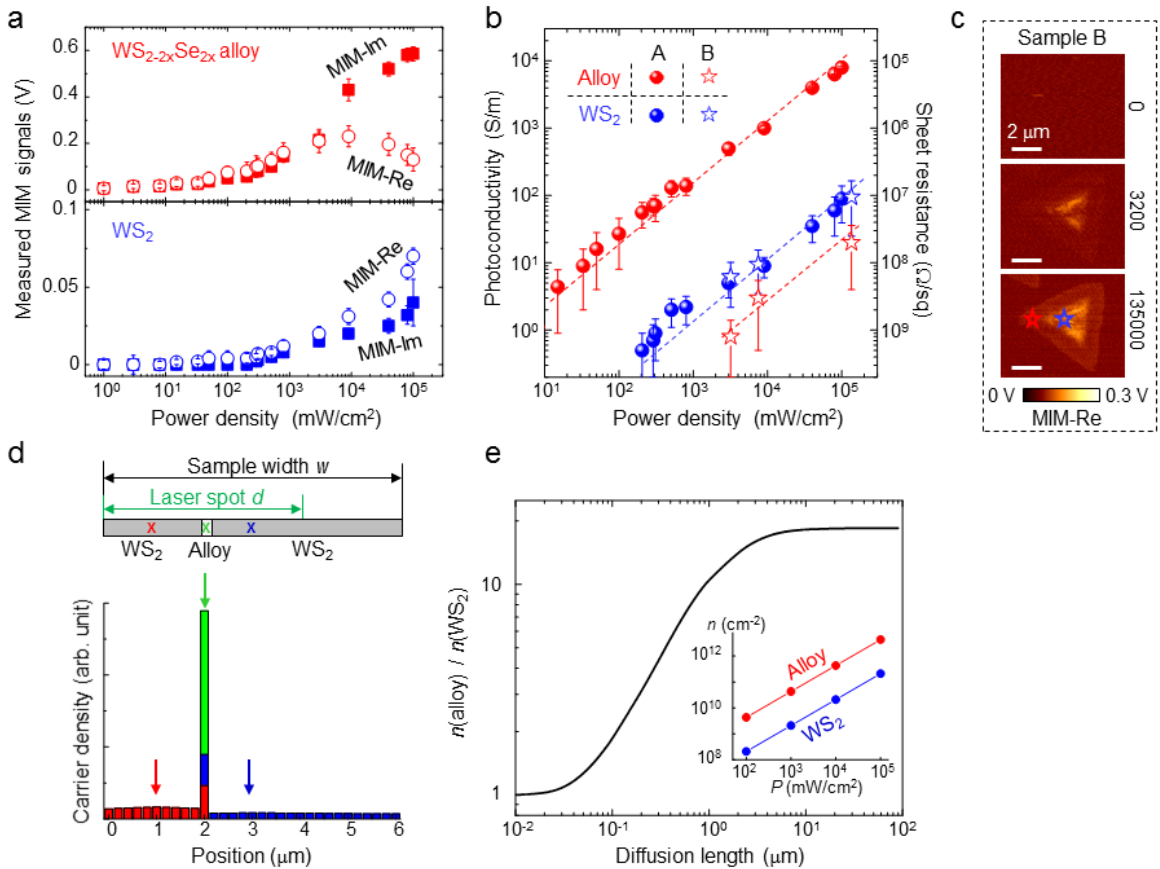

**Figure 3: Power dependence of MIM data and Finite Element Analysis (FEA) of the carrier confinement effect.** a) Average MIM signals as a function of the 532-nm laser intensity on the alloy (top) and WS$_2$ domains (bottom) in sample A. b) Power-dependent photoconductivity for Sample A (solid dots) and Sample B (empty stars). c) MIM-Re images of the single-junction WS$_2$/WS$_{2(1-x)}$Se$_{2x}$ Sample B. Notice the low MIM signal at the alloy domain, suggesting that the enhanced photoconductivity in the multi-junction Sample A is not due to the intrinsic material property of the alloy. All scale bars are 2 μm. d) (top) Schematic of the laser spot and the WS$_2$/alloy/WS$_2$ multi-junction sample. The red, green, and blue crosses are three locations on the left, middle, and right sides of the alloy region, respectively. (bottom) Carrier distribution after diffusion from the three elements denoted above. The charge accumulation in the alloy region is clearly seen in the plot. e) The ratio of $n_{\text{alloy}}/n_{\text{WS2}}$ as a function of the free-carrier diffusion length $L$. Note that the ratio saturates when $L$ is greater than ~ 10 μm. The inset shows that the saturated $n_{\text{alloy}}$ and $n_{\text{WS2}}$ at large $L$ both scale with the laser intensity.

15